\documentclass[twocolumn,showkeys,aps,prb,showpacs]{revtex4-1}
\usepackage{graphicx}
\usepackage[CJKbookmarks,dvipdfm,colorlinks,linkcolor=black,citecolor=black]{hyperref}

\begin{document}

\title{Intrinsic piezoelectric ferromagnetism with large out-of-plane piezoelectric response  in Janus monolayer  $\mathrm{CrBr_{1.5}I_{1.5}}$}

\author{San-Dong Guo$^{1}$, Xiao-Shu Guo$^{1}$,  Xiu-Xia Cai$^{1}$, Wen-Qi Mu$^{1}$  and  Wen-Cai Ren$^{2,3}$}
\affiliation{$^1$School of Electronic Engineering, Xi'an University of Posts and Telecommunications, Xi'an 710121, China}
\affiliation{$^2$Shenyang National Laboratory for Materials Science, Institute of Metal Research,
Chinese Academy of Science, 110016 Shenyang, Liaoning, P. R. China}
\affiliation{$^3$School of Materials Science and Engineering, University of Science and Technology of China,
Shenyang 110016, P. R. China}
\begin{abstract}
A  two-dimensional (2D) material system with both piezoelectricity and ferromagnetic (FM) order, referred to as a 2D piezoelectric ferromagnetism (PFM), may open up unprecedented opportunities for intriguing physics. Inspired by  experimentally synthesized Janus monolayer MoSSe from  $\mathrm{MoS_2}$,
 in this work, the  Janus monolayer  $\mathrm{CrBr_{1.5}I_{1.5}}$ with  dynamic, mechanical and thermal  stabilities is predicted, which is constructed from  synthesized ferromagnetic  $\mathrm{CrI_3}$ monolayer by replacing the top I atomic layer  with Br atoms.
Calculated results show that monolayer  $\mathrm{CrBr_{1.5}I_{1.5}}$ is an  intrinsic FM half semiconductor with  valence and conduction bands being fully spin-polarized in the same spin direction. Furthermore, monolayer  $\mathrm{CrBr_{1.5}I_{1.5}}$  possesses a sizable magnetic anisotropy energy (MAE). By symmetry analysis, it is found that both  in-plane and  out-of-plane piezoelectric polarizations can be
induced by a uniaxial strain in the basal plane.  The calculated  in-plane $d_{22}$ value of 0.557 pm/V is small. However, more excitingly, the  out-of-plane $d_{31}$ is as high as  1.138 pm/V, which  is obviously higher compared with ones of  other 2D known materials. The strong out of-plane piezoelectricity is highly desirable for ultrathin piezoelectric devices.  Moreover, strain engineering is used to tune piezoelectricity of  monolayer  $\mathrm{CrBr_{1.5}I_{1.5}}$. It is found that compressive strain can improve the $d_{22}$, and tensile strain can enhance the $d_{31}$. A FM order  to  antiferromagnetic (AFM) order phase transition can be induced by compressive strain, and the critical point is about 0.95 strain. That is to say that a 2D piezoelectric antiferromagnetism (PAFM) can be achieved by compressive strain, and the corresponding  $d_{22}$ and $d_{31}$ are 0.677 pm/V  and 0.999 pm/V at 0.94 strain, respectively. It is also found that magnetic order has important effects on piezoelectricity of monolayer  $\mathrm{CrBr_{1.5}I_{1.5}}$.
Finally, similar to $\mathrm{CrBr_{1.5}I_{1.5}}$, the PFM can also be realized  in the  monolayer $\mathrm{CrF_{1.5}I_{1.5}}$ and $\mathrm{CrCl_{1.5}I_{1.5}}$. Amazingly, their  $d_{31}$ can reach up to 2.578 pm/V and 1.804 pm/V for  monolayer $\mathrm{CrF_{1.5}I_{1.5}}$ and $\mathrm{CrCl_{1.5}I_{1.5}}$.  Our works propose a realistic way to achieve PFM with large $d_{31}$, making these systems very promising for  multifunctional semiconductor spintronic applications.

\end{abstract}

\pacs{71.20.-b, 77.65.-j, 72.15.Jf, 78.67.-n ~~~~~~~~~~~~~~~~~~~~~~~~~~~~~~~~~~~Email:sandongyuwang@163.com}

\maketitle

\section{Introduction}
The piezoelectric effect is an intrinsic electromechanical coupling in semiconductors with
crystal structures lacking  inversion symmetry.  The reduction in dimensionality of 2D
materials often can  eliminate  inversion symmetry, which
allows them  to be piezoelectric.  Experimentally, the  piezoelectricity  of  $\mathrm{MoS_2}$\cite{q5,q6}, MoSSe\cite{q8}  and $\mathrm{In_2Se_3}$\cite{q8-1} monolayers  have been discovered,  which  pushes the development of  piezoelectric properties of  2D materials.
In theory,  many kinds of 2D
materials have been  predicted to be piezoelectric by density functional theory (DFT) calculations\cite{q7-0,q7-1,q7-2,q7-3,q7-4,q7-5,q7-6,q7-7}.
 The strain-tuned piezoelectric response  has also  been investigated by DFT calculations, and it is proved that strain can improve the  piezoelectric strain  coefficients\cite{q7-8,q7-9,q7-10}.

Great advances have been made on 2D piezoelectric  materials.
However, there are two main issues of 2D piezoelectric materials. One is that  most 2D materials possess solely piezoelectricity.  The multifunctional 2D materials, such as combination of piezoelectricity with   topological
insulating phase or  ferromagnetism,  are of particular interest, whose
exploitation may promise novel device applications. The coexistence of intrinsic piezoelectricity and ferromagnetism has been predicted in 2D vanadium dichalcogenides and   $\mathrm{VSi_2P_4}$\cite{qt1,q15}. The piezoelectric quantum spin Hall insulators (PQSHI) have also been  achieved in monolayer  InXO (X=Se and Te)\cite{gsd1} and Janus monolayer $\mathrm{SrAlGaSe_4}$\cite{gsd2}. Another is that the out-of-plane piezoelectricity in  known 2D materials is absent or  weak.   The strong
out-of-plane piezoelectric effect and its inverse effect are highly desirable for piezoelectric  devices, which is compatible with the bottom/top gate technologies. Many strategies have been made for searching 2D piezoelectric materials
with large $d_{31}$ or $d_{32}$\cite{q7-0,q7-6,q9-0,q9-1,q9}.  A significant improvement is that  the piezoelectric strain coefficient $d_{31}$ of $\mathrm{Sc_2CO_2}$
MXene is up to 0.78 pm/V\cite{q9}.
\begin{figure*}
  \includegraphics[width=10.0cm]{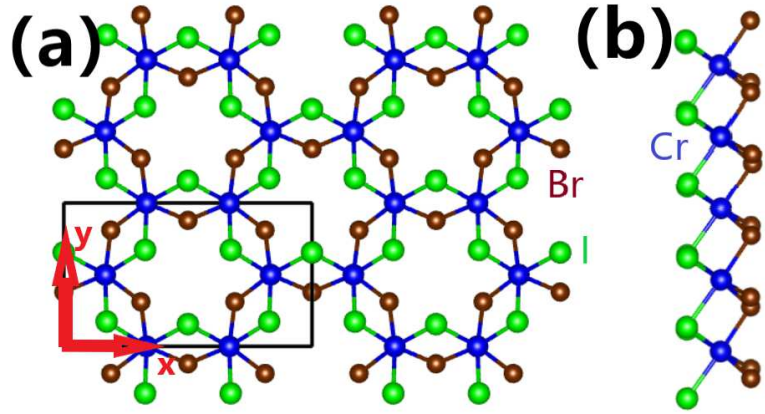}
  \caption{(Color online)The (a) top view and (b) side view of  crystal structure of Janus monolayer  $\mathrm{CrBr_{1.5}I_{1.5}}$.  The  rectangle supercell is marked by  black frame, which is used to calculate the piezoelectric   stress  coefficients. The rectangle's width and height are defined as x and y directions, respectively.}\label{t0}
\end{figure*}
\begin{figure}
  \includegraphics[width=8cm]{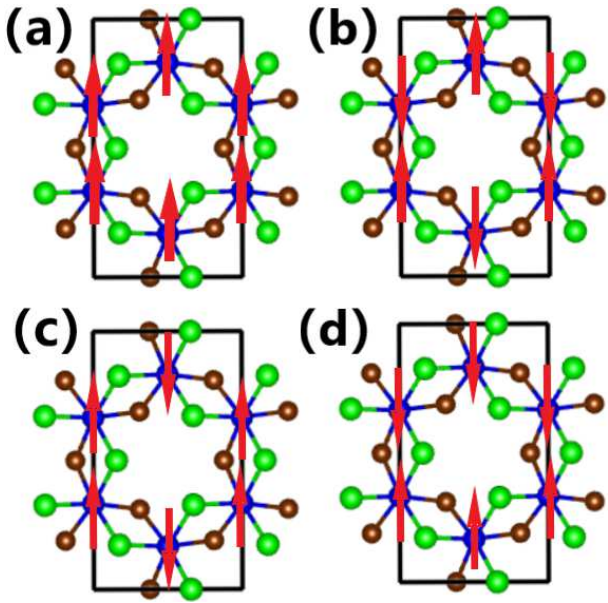}
  \caption{(Color online)The four considered  magnetic configuration of Janus monolayer  $\mathrm{CrBr_{1.5}I_{1.5}}$: FM (a), AF-N$\acute{e}$el (b), AF-zigzag (c), and AF-stripy ordered (d). The  crystal cells
used in the calculations are marked with red arrows as the spin direction of Cr atoms.  }\label{t0-1}
\end{figure}

\begin{figure}
  \includegraphics[width=8cm]{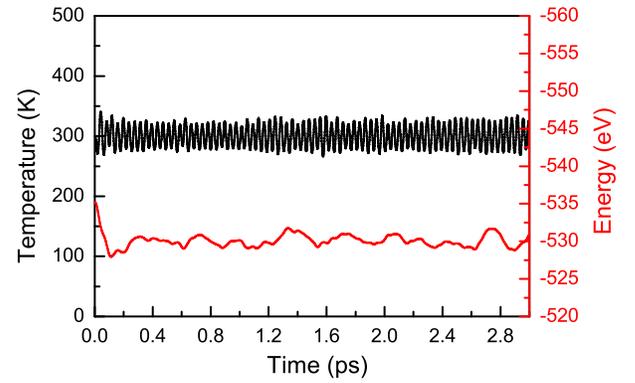}
\caption{(Color online) The temperature and total energy fluctuations of  Janus monolayer  $\mathrm{CrBr_{1.5}I_{1.5}}$  with FM  magnetic configuration at 300 K. }\label{md}
\end{figure}

A natural idea is to search for multifunctional 2D piezoelectric materials with large out-of-plane piezoelectricity. A few types of 2D magnetic materials have been studied\cite{m7-0,m7-1,m7-2,m7-3,m7-3-1,m7-4,m7-5,m7-6,m7-7}. For example, the monolayer $\mathrm{Cr_2Ge_2Te_6}$,  $\mathrm{VS_2}$ and $\mathrm{VSe_2}$ have  been experimentally proved to magnetic materials\cite{m7-2,m7-4}.
The Dirac spin-gapless  semiconductor (SGS) with 100\% spin polarization, high Fermi velocities and  high
Curie temperatures  has been predicted in  $\mathrm{Mn_2C_6Se_{12}}$ and $\mathrm{Mn_2C_6S_6Se_6}$ monolayers\cite{m7-5}.
The 2D high-temperature ferromagnetic half-metal (FMHM) can be realized in transition-metal embedded carbon nitride monolayers\cite{m7-3-1}.
The $\mathrm{CrI_3}$ monolayer  is firstly predicted to be FM order by the first-principle calculations\cite{m7-8}, and then is confirmed experimentally\cite{m7-6}. A series of studies have been carried out to
explore the magnetic related properties in $\mathrm{CrI_3}$ monolayer\cite{m8-0,m8-1,m8-2,m8-3,m8-4,m8-5}. These provide many new  possibilities to combine the piezoelectricity and magnetism into the same kind of 2D material.

It is noted that the $\mathrm{CrI_3}$ monolayer has  sandwiched I-Cr-I structure with  inversion symmetry, and then possesses no piezoelectricity. However, it is possible to construct Janus structure based on  $\mathrm{CrI_3}$ monolayer, and then produce piezoelectric effect.
Janus monolayer MoSSe has been synthesized experimentally from $\mathrm{MoS_2}$ monolayer by  breaking the out-of-plane structural symmetry\cite{p1}. That is, the  Janus monolayer MoSSe can be constructed  by  replacing one of two  S   layers with Se  atoms in  $\mathrm{MoS_2}$ monolayer.
In this work,  Janus monolayer  $\mathrm{CrBr_{1.5}I_{1.5}}$ is constructed from  synthesized ferromagnetic  $\mathrm{CrI_3}$ monolayer by replacing the top I atomic layer  with Br atoms, which is  dynamically, mechanically  and thermally stable.  It is found  that monolayer  $\mathrm{CrBr_{1.5}I_{1.5}}$ is an  intrinsic FM half semiconductor with a sizable MAE.  Although the calculated  in-plane $d_{22}$ (0.557 pm/V) is small, the  out-of-plane $d_{31}$ (1.138 pm/V)  is very large, which  is obviously higher than  ones of  other 2D known materials. It is proved that strain engineering can effectively tune piezoelectricity of  monolayer  $\mathrm{CrBr_{1.5}I_{1.5}}$.
A 2D PAFM can also  be achieved  in monolayer  $\mathrm{CrBr_{1.5}I_{1.5}}$ by compressive strain, and the calculated results show that magnetic order has important influences  on piezoelectricity of monolayer  $\mathrm{CrBr_{1.5}I_{1.5}}$.
It is also proved that the PFM can also be achieved in monolayer  $\mathrm{CrF_{1.5}I_{1.5}}$ and monolayer  $\mathrm{CrCl_{1.5}I_{1.5}}$, which show very large $d_{31}$ of 2.578 pm/V and 1.804 pm/V.

The rest of the paper is organized as follows. In the next
section, we shall give our computational details and methods.
 In  the next few sections,  we shall present structural stabilities, electronic structures, and piezoelectric properties of monolayer  $\mathrm{CrBr_{1.5}I_{1.5}}$, along with strain effects on its piezoelectric properties. Finally, we shall give our discussion and conclusions.

\begin{figure}
  \includegraphics[width=8cm]{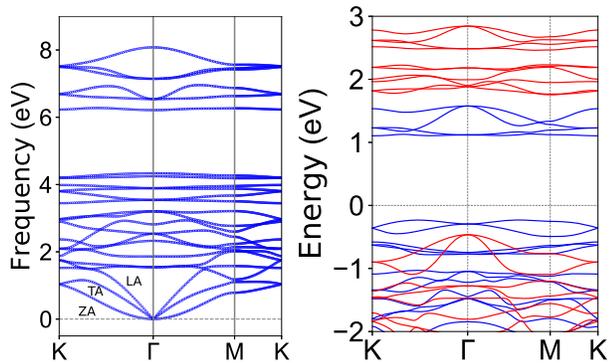}
  \caption{(Color online)Left: the phonon band dispersions  of Janus monolayer  $\mathrm{CrBr_{1.5}I_{1.5}}$  with FM  magnetic configuration. Right: the  energy band structures of $\mathrm{CrBr_{1.5}I_{1.5}}$ with FM state. The blue (red) lines represent the band
structure in the spin-up (spin-down) direction.}\label{band}
\end{figure}

\begin{figure}
  \includegraphics[width=8cm]{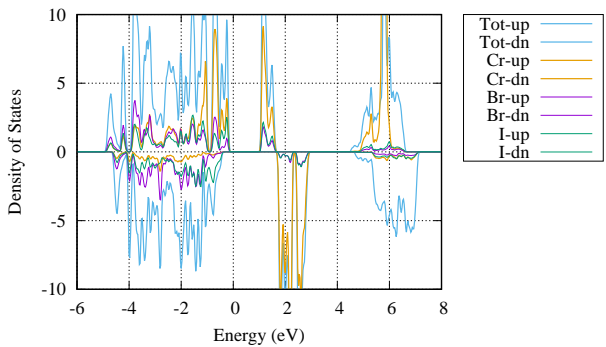}
  \caption{(Color online) Calculated total and atomic partial density of states of Janus monolayer  $\mathrm{CrBr_{1.5}I_{1.5}}$  with FM  magnetic configuration.  }\label{dos}
\end{figure}
\begin{figure}
  \includegraphics[width=8cm]{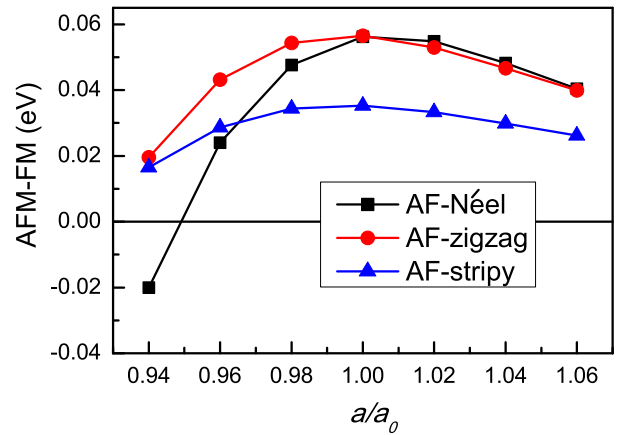}
  \caption{(Color online) Calculated energy differences of   AF-N$\acute{e}$el, AF-zigzag and AF-stripy  with respect to  FM state as a function of strain with rectangle supercell.  }\label{energy}
\end{figure}

\section{Computational detail}
Within DFT\cite{1}, we perform the main calculations  with spin-polarization using the the plane-wave code VASP\cite{pv1,pv2,pv3} within the projector augmented-wave (PAW) method.  The popular  generalized gradient approximation of Perdew, Burke and  Ernzerhof  (GGA-PBE)\cite{pbe} is used as the exchange-correlation  functional.
The kinetic energy cutoff is set to 500 eV with the total energy  convergence criterion for $10^{-8}$ eV.  All the lattice constants and atomic coordinates are optimized until the
force on each atom is less than 0.0001 $\mathrm{eV.{\AA}^{-1}}$.  A vacuum spacing of more than 18 $\mathrm{{\AA}}$ is used  to avoid interactions between two neighboring images.
The elastic stiffness tensor  $C_{ij}$  and piezoelectric stress tensor $e_{ij}$  are carried out by using strain-stress relationship (SSR) and   density functional perturbation theory (DFPT) method\cite{pv6}, respectively.
A Monkhorst-Pack k-mesh of 8$\times$8$\times$1 is used to sample the Brillouin Zone (BZ) for the  calculations of electronic structure  and elastic coefficients $C_{ij}$ , and  a mesh of 4$\times$8$\times$1 k-points
for  the energy of different magnetic configurations and piezoelectric   stress  coefficients $e_{ij}$.
The 2D elastic coefficients $C^{2D}_{ij}$
 and   piezoelectric stress coefficients $e^{2D}_{ij}$
have been renormalized by   $C^{2D}_{ij}$=$Lz$$C^{3D}_{ij}$ and $e^{2D}_{ij}$=$Lz$$e^{3D}_{ij}$, where the $Lz$ is  the length of unit cell along z direction.  By finite displacement method, the interatomic force constants (IFCs) are obtained based on
the 4$\times$4$\times$1 supercell with FM ground state. Based on the harmonic IFCs, the
phonon dispersions are evaluated
using Phonopy code\cite{pv5}.

\begin{figure*}
  \includegraphics[width=12cm]{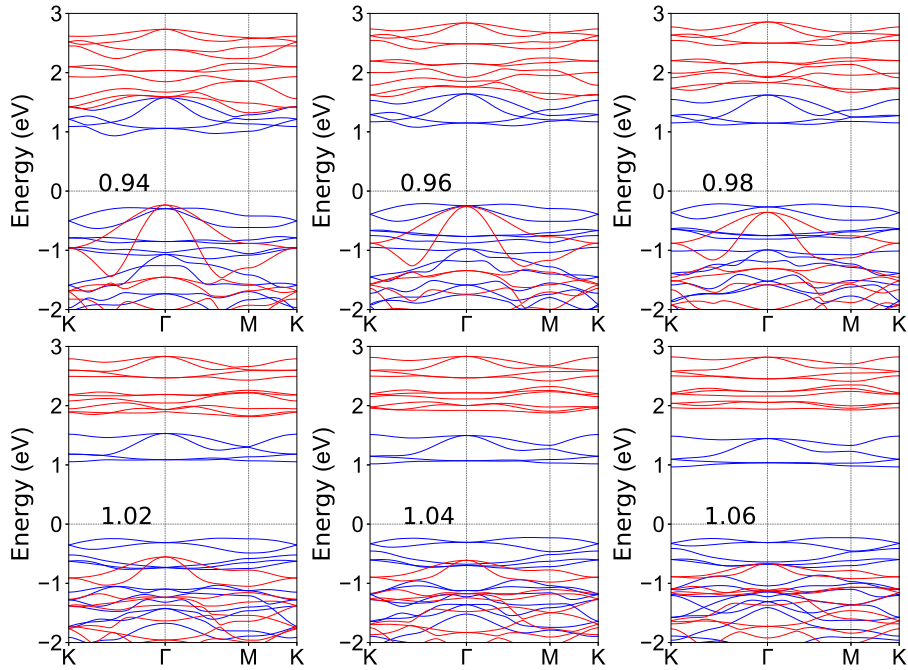}
\caption{(Color online) The energy band structures  of Janus monolayer  $\mathrm{CrBr_{1.5}I_{1.5}}$  with FM  magnetic configuration with $a/a_0$ changing from 0.94 to 1.06.}\label{t3}
\end{figure*}
\begin{figure}
   \includegraphics[width=8.0cm]{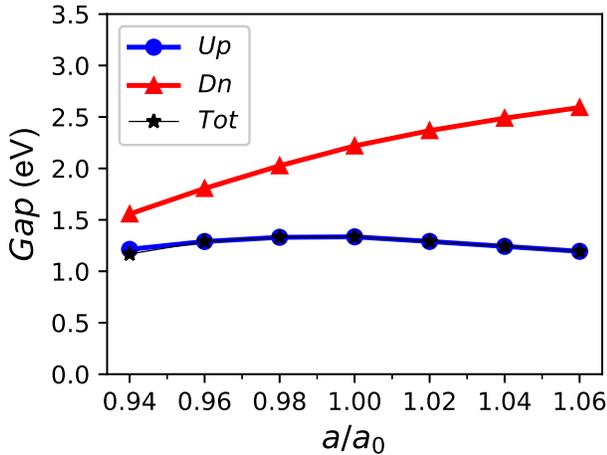}
  \caption{(Color online) The majority-spin gap ($Up$), the minority-spin gap ($Dn$) and the total gap ($Tot$) of FM $\mathrm{CrBr_{1.5}I_{1.5}}$ as a function of  strain $a/a_0$.}\label{t4-gap}
\end{figure}

\section{Structure and stability}
The structure of Janus monolayer  $\mathrm{CrBr_{1.5}I_{1.5}}$ is similar to monolayer $\mathrm{CrI_3}$  monolayer, which contains three atomic sublayers with Cr layer sandwiched between Br and I layers. It is well known that Janus transition metal dichalchogenides (TMD) Monolayer MoSSe  has been synthesized by replacing the top S atomic layer in  $\mathrm{MoS_2}$ with Se atoms\cite{p1}.  Using  the same idea, the  Janus monolayer   $\mathrm{CrBr_{1.5}I_{1.5}}$ can be constructed  by  replacing one of two I  layers with Br  atoms in monolayer  $\mathrm{CrI_3}$.
The schematic crystal structures of Janus monolayer  $\mathrm{CrBr_{1.5}I_{1.5}}$ are shown in \autoref{t0}.
The monolayer  $\mathrm{CrI_3}$ has centrosymmetry with  $\bar{3}m$ point-group symmetry (No.162), but
monolayer  $\mathrm{CrBr_{1.5}I_{1.5}}$ loses centrosymmetry and horizontal mirror symmetry with  $3m$ point-group symmetry (No.157), which will induce both in-plane and out-of-plane piezoelectricity.

 Four  different magnetic configurations (\autoref{t0-1}) are considered to  evaluate the magnetic ground state of monolayer  $\mathrm{CrBr_{1.5}I_{1.5}}$, which are used to investigate the ground state of  monolayer  $\mathrm{CrX_3}$ (X=F, Cl, Br and I)\cite{m7-8}.  The energy of  AF-N$\acute{e}$el, AF-zigzag,  AF-stripy state and non-magnetic (NM) state with respect to FM state are
56.1 meV, 56.4 meV, 35.2 meV and 6.802 eV with rectangle supercell. Our calculated results show that the FM order is the most stable magnetic state. This means that
ferromagnetism in monolayer  $\mathrm{CrI_3}$ is retained by elements substitution to construct Janus structure.
 The optimized lattice constants with FM state is 6.744 $\mathrm{{\AA}}$, which falls between those of the
$\mathrm{CrBr_3}$ (6.433 $\mathrm{{\AA}}$) and $\mathrm{CrI_3}$ (7.008 $\mathrm{{\AA}}$) monolayers\cite{m7-8}.
For monolayer  $\mathrm{CrBr_{1.5}I_{1.5}}$,
the difference in atomic sizes and electronegativities of Br and I atoms leads to inequivalent Cr-Br and Cr-I bond lengths (Br-Cr-Br and I-Cr-I bond  angles), and they are
2.542 $\mathrm{{\AA}}$ and 2.719 $\mathrm{{\AA}}$ (92.036 and 89.432), which can induce a built-in electric field.

The ab initio molecular dynamics (AIMD) simulations  using NVT ensemble are performed to
assess the thermal stability of the monolayer $\mathrm{CrBr_{1.5}I_{1.5}}$ at room
temperature.  \autoref{md} shows the temperature and total energy fluctuations of $\mathrm{CrBr_{1.5}I_{1.5}}$ monolayer as a function of the simulation time.  Calculated results show no obvious structural disruption with the temperature and total energy
 fluctuates being small at the end of the MD simulation at 300 K, which  confirms the thermodynamical stability of the $\mathrm{CrBr_{1.5}I_{1.5}}$ monolayer at room temperature.

The dynamical stability of the  $\mathrm{CrBr_{1.5}I_{1.5}}$ monolayer is analyzed
by  the phonon spectra, which is plotted in \autoref{band}. There are  twenty-one optical and three acoustical phonon
branches with a total of twenty-four branches due to eight
atoms per cell.
The longitudinal acoustic (LA) and transverse
acoustic (TA) modes mean in-plane vibrations, while the
ZA branch represents the out-of-plane vibrations. It is clearly seen  that
the ZA branch is   quadratic near the zone center, as typical characteristics of 2D materials\cite{r1,r2}.
 All phonon frequencies are positive,
confirming the dynamical stability of $\mathrm{CrBr_{1.5}I_{1.5}}$ monolayer, which means  that it can exist as a free-standing
2D crystal.

It is important to check
the mechanical stability of $\mathrm{CrBr_{1.5}I_{1.5}}$ monolayer for practical application. Therefore, we calculate elastic
constants  using the SSR method. Using Voigt notation, the elastic tensor with $\bar{3}m$ point-group symmetry for 2D materials can be expressed as:
\begin{equation}\label{pe1-4}
   C=\left(
    \begin{array}{ccc}
      C_{11} & C_{12} & 0 \\
     C_{12} & C_{11} &0 \\
      0 & 0 & (C_{11}-C_{12})/2 \\
    \end{array}
  \right)
\end{equation}
The calculated $C_{11}$ and $C_{12}$ are 29.75 $\mathrm{Nm^{-1}}$ and 8.26 $\mathrm{Nm^{-1}}$, which are between ones of $\mathrm{CrBr_3}$  and $\mathrm{CrI_3}$ monolayers\cite{m7-8}.  The calculated $C_{11}$$>$0 and $C_{11}-C_{12}$$>$0 satisfy the  Born  criteria of mechanical stability\cite{ela},  confirming the mechanical stability of $\mathrm{CrBr_{1.5}I_{1.5}}$ monolayer.   We also calculate the Young's moduli $C_{2D}$,  shear modulus $G_{2D}$ and Poisson's ratio $\nu$ using the method suggested by Andrew et al, and they are 27.46  $\mathrm{Nm^{-1}}$, 10.75  $\mathrm{Nm^{-1}}$ and  0.278, respectively.
These indicate  that  monolayer  $\mathrm{CrBr_{1.5}I_{1.5}}$  can be easily tuned by strain, which is favorable for novel flexible piezotronics and nanoelectronics.

\begin{figure}
  \includegraphics[width=8cm]{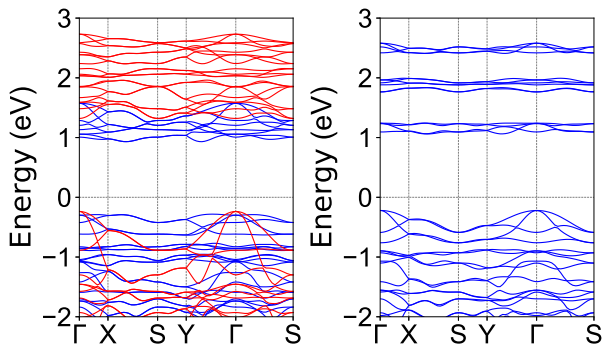}
  \caption{(Color online)The  energy band structures of $\mathrm{CrBr_{1.5}I_{1.5}}$ with FM (Left) and AF-N$\acute{e}$el (Right) states at 0.94 strain. The blue (red) lines represent the band
structure in the spin-up (spin-down) direction.}\label{t5}
\end{figure}

\begin{figure}
   \includegraphics[width=8cm]{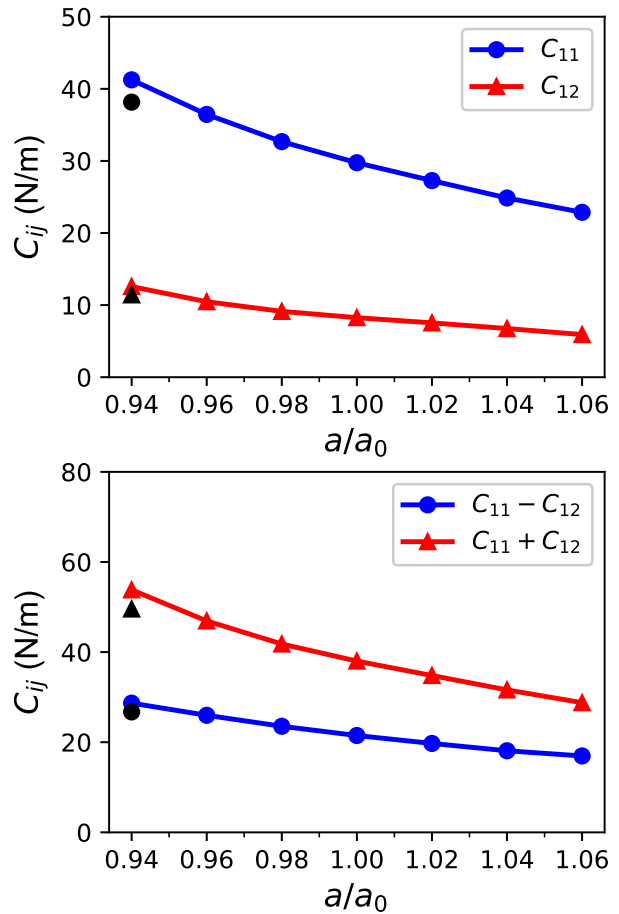}
  \caption{(Color online) For Janus monolayer  $\mathrm{CrBr_{1.5}I_{1.5}}$ with FM state,  the elastic constants  $C_{ij}$ with the application of  biaxial strain (0.94 to 1.06). The black marks mean AF-N$\acute{e}$el results at 0.94 strain.}\label{cc}
\end{figure}

\begin{figure}
   \includegraphics[width=8cm]{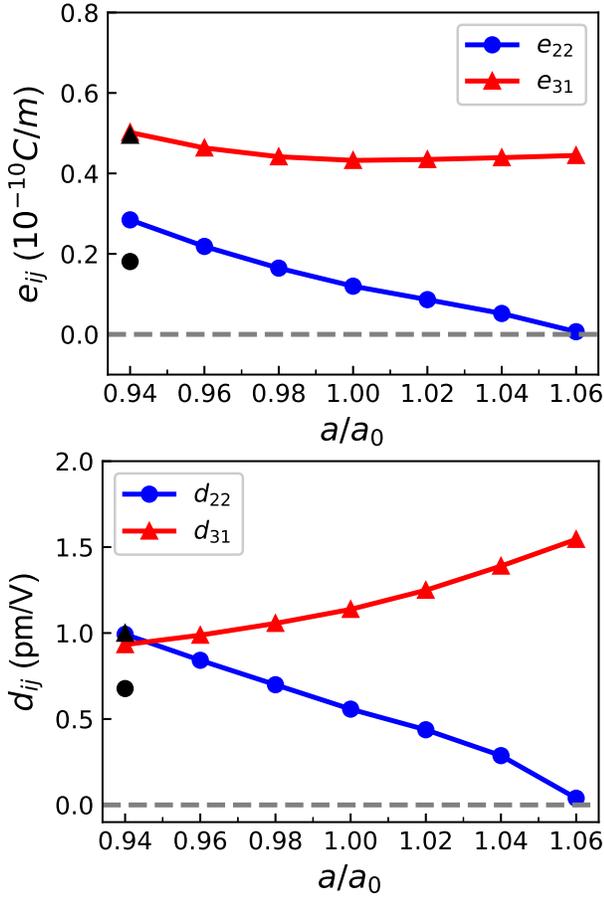}
    \caption{(Color online) For Janus monolayer  $\mathrm{CrBr_{1.5}I_{1.5}}$ with FM state, the piezoelectric stress coefficients  ($e_{22}$ and  $e_{31}$) and the piezoelectric strain coefficients  ($d_{22}$ and $d_{31}$)  with the application of  biaxial strain (0.94 to 1.06). The black marks mean AF-N$\acute{e}$el results at 0.94 strain. }\label{ed}
\end{figure}

\section{Electronic structure}
To exhibit  piezoelectricity, the monolayer $\mathrm{CrBr_{1.5}I_{1.5}}$  not only  should   lack inversion symmetry,  but also should be a semiconductor. So, we  investigate the electronic structures of $\mathrm{CrBr_{1.5}I_{1.5}}$ monolayer with FM ground state, and
the energy bands and  atomic  partial density of states (DOS) are plotted in \autoref{band} and \autoref{dos}, respectively.
It is found that $\mathrm{CrBr_{1.5}I_{1.5}}$ monolayer is an indirect gap semiconductor with gap value of 1.335 eV.
 Moreover,  the valence and conduction bands  near the Fermi level are exclusively contributed by the same spin-up
component,  showing a typical half-semiconductor character. The difference of the band edge energy between the two spin
components  for the conduction band minimum (CBM) and the valance band
maximum (VBM) are 0.884 eV and 0.233 eV, respectively. According to DOS,  the occupied Cr-$3d$ orbitals are mainly
found in the spin-up direction,  and the spin-down Cr-$3d$ states
are almost  unoccupied. For both
spin directions, the conduction band is dominated by Cr-$3d$ states, which are weakly
hybridized with the Br-$3p$ and I-$3p$ states. In the spin-down direction, the valence band  are  almost pure Br-$3p$ and I-$3p$ character. For the spin-up direction of the valence
band, the states are contributed by the Br-$3p$ and I-$3p$ states with a
mixture of Cr-$3d$ states. In fact, many electronic properties of monolayer $\mathrm{CrBr_{1.5}I_{1.5}}$ are similar to ones of $\mathrm{CrI_3}$ monolayer\cite{m7-8}.

 The magnetic moment of primitive cell is equal to 6 $\mu_B$ accurately, which is consistent with its semiconducting property.
The local magnetic moments of Cr is 2.985 $\mu_B$, which  suggests
that monolayer $\mathrm{CrBr_{1.5}I_{1.5}}$ is robust intrinsic ferromagnetic 2D semiconductor with large magnetic moments.
MAE is an important parameter to confirm ferromagnetic behavior of monolayer $\mathrm{CrBr_{1.5}I_{1.5}}$.
The  small MAE will result in superparamagnetic rather than ferromagnetic behavior. By using GGA+spin orbital coupling (SOC), it is found  that an easy axis
is along the c-direction for monolayer $\mathrm{CrBr_{1.5}I_{1.5}}$, and the corresponding MAE is  356 $\mu$eV per Cr atom.
For $\mathrm{CrBr_{3}}$ and  $\mathrm{CrI_{3}}$ monolayers, the easy axis
is also along the c-direction, and the MAE of monolayer $\mathrm{CrBr_{1.5}I_{1.5}}$ is between  ones of them (185.5 $\mu$eV per Cr atom for $\mathrm{CrBr_{3}}$ and  685.5 $\mu$eV per Cr atom for $\mathrm{CrI_{3}}$)\cite{m7-8}.

\begin{figure}
    \includegraphics[width=8cm]{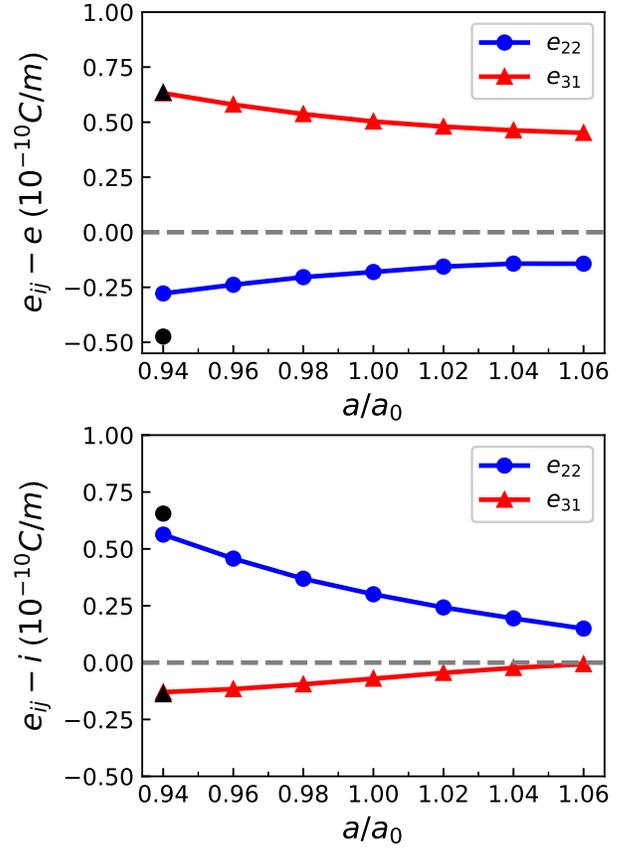}
\caption{(Color online) For Janus monolayer  $\mathrm{CrBr_{1.5}I_{1.5}}$ with FM state, the ionic contribution and electronic contribution to  $e_{22}$ and  $e_{31}$ with the application of  biaxial strain (0.94 to 1.06). The black marks mean AF-N$\acute{e}$el results at 0.94 strain. }\label{ed1}
\end{figure}

\section{Piezoelectric properties}
The $\mathrm{CrI_3}$ monolayer  with  $\bar{3}m$ point-group symmetry  are centrosymmetric,  showing no piezoelectricity.
 The  $\mathrm{CrBr_{1.5}I_{1.5}}$ monolayer with $3m$ point-group symmetry lacks  both inversion symmetry and reflectional
symmetry across the xy plane, which means that   both $e_{22}$/$d_{22}$ and $e_{31}$/$d_{31}$ with  defined x and y direction in \autoref{t0} are nonzero. For 2D materials, only the in-plane strain and stress are taken into account\cite{q7-0,q7-1,q7-2,q7-3,q7-4,q7-5,q7-6,q7-7}, and  the  piezoelectric stress   and strain tensors by using  Voigt notation  can become:
 \begin{equation}\label{pe1-1}
 e=\left(
    \begin{array}{ccc}
     0 & 0 & -e_{22} \\
     -e_{22} & e_{22} & 0 \\
      e_{31} & e_{31} & 0 \\
    \end{array}
  \right)
    \end{equation}

  \begin{equation}\label{pe1-2}
  d= \left(
    \begin{array}{ccc}
      0 & 0 & -2d_{22} \\
       -d_{22} & d_{22} & 0 \\
      d_{31} & d_{31} &0 \\
    \end{array}
  \right)
\end{equation}
With a applied uniaxial in-plane strain,   both in-plane and vertical piezoelectric polarization ($e_{22}$/$d_{22}$$\neq$0 and $e_{31}$/$d_{31}$$\neq$0) can be produced. However, by  imposing biaxial in-plane strain, the
out-of-plane one still will remain , while the
in-plane piezoelectric response will be suppressed($e_{11}$/$d_{11}$=0 and $e_{31}$/$d_{31}$$\neq$0).
The independent $d_{22}$ and $d_{31}$ are can be attained by $e_{ik}=d_{ij}C_{jk}$:
\begin{equation}\label{pe2}
    d_{22}=\frac{e_{22}}{C_{11}-C_{12}}~~~and~~~d_{31}=\frac{e_{31}}{C_{11}+C_{12}}
\end{equation}
\begin{table*}
\centering \caption{For $\mathrm{CrX_{1.5}I_{1.5}}$(X=F, Cl and Br) monolayers, the lattice constants $a_0$ ($\mathrm{{\AA}}$),  the elastic constants $C_{ij}$ in $\mathrm{Nm^{-1}}$, the piezoelectric stress coefficients $e_{ij}$ in $10^{-10}$ C/m, the piezoelectric strain coefficients $d_{ij}$ in pm/V, MAE in $\mu eV$/Cr and easy axis (EA). }\label{tab0}
  \begin{tabular*}{0.96\textwidth}{@{\extracolsep{\fill}}cccccccccc}
  \hline\hline
Name&$a_0$&  $C_{11}$& $C_{12}$&$e_{22}$ & $e_{31}$ & $d_{22}$ &$d_{31}$ &MAE &EA\\\hline
$\mathrm{CrF_{1.5}I_{1.5}}$& 6.250&49.97&16.34&1.339&1.710&3.983&2.578&2151& $ab$ \\\hline
$\mathrm{CrCl_{1.5}I_{1.5}}$&6.590&34.88&9.94&0.238&0.809&0.956&1.804&110&$c$ \\\hline
$\mathrm{CrBr_{1.5}I_{1.5}}$&6.744&29.75&8.26&0.119&0.432&0.557&1.138&356&$c$ \\\hline\hline
\end{tabular*}
\end{table*}

The orthorhombic supercell is used  as the
computational unit cell (in \autoref{t0}) to calculate the  $e_{ij}$  of $\mathrm{CrBr_{1.5}I_{1.5}}$ monolayer.
The calculated $e_{22}$ is 0.119$\times$$10^{-10}$ C/m  with ionic part 0.300$\times$$10^{-10}$ C/m  and electronic part -0.181$\times$$10^{-10}$ C/m, and $e_{31}$ for 0.432$\times$$10^{-10}$ C/m with ionic contribution -0.071$\times$$10^{-10}$ C/m  and electronic contribution 0.503$\times$$10^{-10}$ C/m. It is found that the electronic and
ionic polarizations  have  opposite signs for both $e_{22}$ and $e_{31}$.  The ionic contribution
to the in-plane piezoelectricity is larger than the electronic
contribution. However,  the electronic contributions dominate the out-of-plane piezoelectricity.
Based on \autoref{pe2}, the $d_{22}$  and $d_{31}$ can be attained from previous calculated $C_{ij}$ and $e_{ij}$.
The calculated  $d_{22}$  and $d_{31}$ are 0.557 pm/V and 1.138 pm/V.   A large out-of-plane piezoelectric response is
highly desired for 2D materials, which is compatible with the
nowadays bottom/top gate technologies.   The $d_{31}$ of $\mathrm{CrBr_{1.5}I_{1.5}}$ monolayer is obviously higher compared with ones of other 2D known materials, including the oxygen functionalized MXenes (0.40-0.78 pm/V)\cite{q9}, Janus TMD monolayers (0.03 pm/V)\cite{q7-0},
functionalized h-BN (0.13 pm/V)\cite{o1}, kalium decorated graphene (0.3
pm/V)\cite{o2}, Janus group-III materials (0.46 pm/V)\cite{q7-6}, Janus BiTeI/SbTeI  monolayer (0.37-0.66 pm/V)\cite{o3} and $\alpha$-$\mathrm{In_2Se_3}$
(0.415 pm/V)\cite{o4}.
To the
best of our knowledge, the $d_{31}$ is less than one among all pure  2D materials. So, it is  very peculiar that the $d_{31}$ of $\mathrm{CrBr_{1.5}I_{1.5}}$ monolayer is as high as 1.138 pm/V.

\begin{figure}
  \includegraphics[width=8cm]{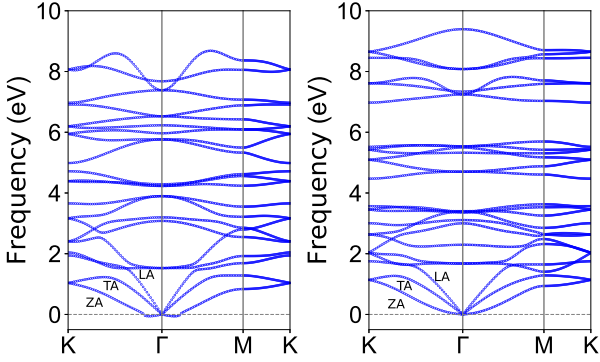}
  \caption{(Color online)the phonon band dispersions  of Janus monolayer  $\mathrm{CrF_{1.5}I_{1.5}}$ (Left) and $\mathrm{CrCl_{1.5}I_{1.5}}$ (Right) with FM  magnetic configuration.}\label{ph11}
\end{figure}

\section{Strain effects}
The strain can effectively tune the electronic structures and piezoelectric properties of 2D materials\cite{q7-8,q7-9,q7-10}.
Here, we use $a/a_0$ to simulate the biaxial strain, where  $a$ and $a_0$ are  the strained and  unstrained lattice constants, respectively.
To determine the ground state of strained   $\mathrm{CrF_{1.5}I_{1.5}}$ monolayer, four  different initial magnetic configurations (\autoref{t0-1}) are considered.  The energy differences of   AF-N$\acute{e}$el, AF-zigzag and AF-stripy  with respect to  FM state as a function of strain with rectangle supercell are shown in \autoref{energy}.  It is found that a magnetic phase transition can be induced by compressive strain with the critical point being about 0.95, which implies  the robustness of the intrinsic
ferromagnetism in $\mathrm{CrF_{1.5}I_{1.5}}$ monolayer.
 Calculated results show that $\mathrm{CrF_{1.5}I_{1.5}}$ monolayer prefers FM ground state with $a/a_0$ being greater than about 0.95 in considered strain range, and the  AF-N$\acute{e}$el become ground state with $a/a_0$ being less than about 0.95.
Similar phenomenon can also be found in $\mathrm{CrI_3}$ monolayer, and the AF-N$\acute{e}$el phase becomes the most stable phase at 0.92
strain\cite{m7-8}.

The  energy band structures of FM $\mathrm{CrF_{1.5}I_{1.5}}$ monolayer with strain from 0.94 to 1.06 except 1.00 are plotted  in \autoref{t3}, and the majority-spin,  minority-spin  and total gaps  are shown in \autoref{t4-gap}. It is clearly seen that strained $\mathrm{CrF_{1.5}I_{1.5}}$ monolayer  are all indirect gap semiconductors in considered strain range. It is found that the majority-spin and total gaps coincide except 0.94 strain, which means that $\mathrm{CrF_{1.5}I_{1.5}}$ monolayer holds  half-semiconductor character. At 0.94 strain, the VBM is at minority-spin channel from previous majority-spin one.
From 1.06 to 0.94 strain, strain makes both conduction and valence bands of minority-spin channel
move toward  Fermi level, which leads to the reduced minority-spin gap. The majority-spin gap shows a nonmonotonic behavior, which is mainly due to change of CBM.  In fact, at 0.94 strain, the  AF-N$\acute{e}$el becomes ground state, and we plot the energy bands along with FM states in \autoref{t5} using orthorhombic supercell.  The  AF-N$\acute{e}$el state still is an indirect gap semiconductor with the gap value of 1.282 eV, and the local magnetic moments of Cr is 2.893 $\mu_B$.

\begin{figure}
  \includegraphics[width=8cm]{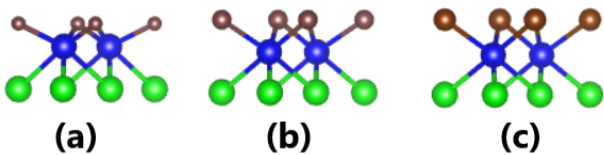}
  \caption{(Color online)The side views of crystal structure  of $\mathrm{CrF_{1.5}I_{1.5}}$ (a),  $\mathrm{CrCl_{1.5}I_{1.5}}$ (b) and $\mathrm{CrBr_{1.5}I_{1.5}}$ (c).}\label{cccc}
\end{figure}
The strain engineering has been proved to be an very effective way  to enhance  piezoelectric properties of 2D materials, and then  the strain effects on piezoelectric properties of  $\mathrm{CrBr_{1.5}I_{1.5}}$ monolayer are performed.
The elastic constants  including $C_{11}$, $C_{12}$,  $C_{11}$-$C_{12}$ and  $C_{11}$+$C_{12}$ of  $\mathrm{CrBr_{1.5}I_{1.5}}$ monolayer  with FM state  as a function of  biaxial  strain are plotted in \autoref{cc}, along with ones of  AF-N$\acute{e}$el state at 0.94 strain.
It is clearly seen that $C_{11}$, $C_{12}$,  $C_{11}$-$C_{12}$ and  $C_{11}$+$C_{12}$ are  all decreases with increasing strain from 0.94 to 1.06 strain.  It is found that $C_{11}$, $C_{12}$,  $C_{11}$-$C_{12}$ and  $C_{11}$+$C_{12}$ with AF-N$\acute{e}$el state are lower than ones with FM state at 0.94 strain. So, it is  important to  consider the magnetic configurations for calculating elastic constants.
 Calculated results show that   the $\mathrm{CrBr_{1.5}I_{1.5}}$ monolayer is mechanically stable in the considered strain range, since the calculated elastic constants satisfy   the mechanical stability criteria\cite{ela}.

 The piezoelectric  stress  coefficients  ($e_{22}$ and $e_{31}$) along  the ionic  and electronic contributions and  piezoelectric  strain  coefficients ($d_{22}$ and $d_{31}$) of $\mathrm{CrBr_{1.5}I_{1.5}}$ monolayer  with FM state as a function of  biaxial  strain are plotted in \autoref{ed} and \autoref{ed1}, along with ones of  AF-N$\acute{e}$el state at 0.94 strain.
It is found that the compressive strain can enhance the $d_{22}$ due to improved $e_{22}$ based on \autoref{pe2}, and the $d_{22}$ improves to 0.993 pm/V at 0.94 strain from unstrained 0.557 pm/V. The tensile strain can decrease the $d_{22}$, and  the $d_{22}$ at 1.06 strain reduces  to 0.039 pm/V due to very small $e_{22}$ (0.0066$\times$$10^{-10}$ C/m).  For $d_{31}$, the opposite strain dependence is observed, and the tensile strain can improve $d_{31}$ due to reduced $C_{11}$+$C_{12}$.  At 1.06 strain, the $d_{31}$ of   $\mathrm{CrBr_{1.5}I_{1.5}}$ monolayer is  1.545 pm/V, increased  by 36\% with respect to unstrained one.  In considered strain range, the electronic and
ionic parts  have  opposite signs for both $e_{22}$ and $e_{31}$, and they  (absolute value) all decreases with strain from 0.94 to 1.06.

The magnetic configuration may have important effects on piezoelectric  coefficients, and a magnetic phase transition may induce the  jump  of  piezoelectric  coefficients. We recalculate the  $e_{22}$ and $e_{31}$ along  the ionic  and electronic contributions and $d_{22}$ and $d_{31}$ of $\mathrm{CrBr_{1.5}I_{1.5}}$ monolayer  with  AF-N$\acute{e}$el state at 0.94 strain.  It is found that magnetic configuration has important effect on $e_{22}$ from  0.285$\times$$10^{-10}$ C/m of  FM state  to 0.181$\times$$10^{-10}$ C/m of AF-N$\acute{e}$el state, and has little influence on $e_{31}$ (0.502$\times$$10^{-10}$ C/m for FM and   0.495$\times$$10^{-10}$ C/m for AF-N$\acute{e}$el).   The similar effects on $d_{22}$ and $d_{31}$ also can be found, and  the $d_{22}$  ($d_{31}$) changes from 0.993 pm/V (0.933 pm/V) of FM state to  0.677 pm/V (0.999 pm/V) of  AF-N$\acute{e}$el state.
It is also found that  magnetic configuration has important effects on both  the electronic and
ionic parts of $e_{22}$, and has neglectful influences on ones of $e_{31}$. So, it is very important to consider magnetic order for piezoelectric  coefficients. It is interesting that the PAFM can be induced by compressive strain, which may open up potential opportunities for intriguing physics and novel devices.

\begin{figure}
  \includegraphics[width=8cm]{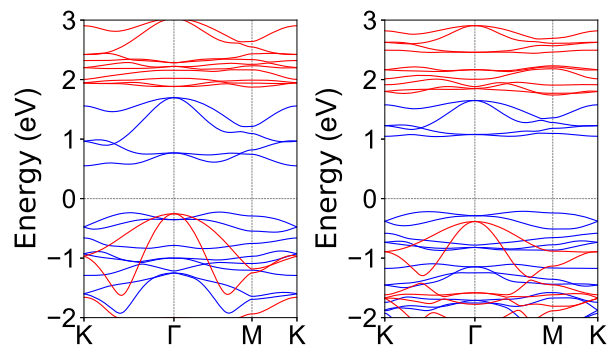}
  \caption{(Color online)The  energy band structures of $\mathrm{CrF_{1.5}I_{1.5}}$ (Left) and $\mathrm{CrCl_{1.5}I_{1.5}}$ (Right) with FM  state. The blue (red) lines represent the band
structure in the spin-up (spin-down) direction.}\label{band0.94}
\end{figure}

\section{Discussion and Conclusion}
In fact, one of two I  layers of  monolayer  $\mathrm{CrI_3}$ can also be replaced by F or Cl atoms, namely monolayer $\mathrm{CrF_{1.5}I_{1.5}}$ and  $\mathrm{CrCl_{1.5}I_{1.5}}$. For $\mathrm{CrCl_{1.5}I_{1.5}}$ monolayer, the FM order still is the ground state by comparing energy difference  of  four  different initial magnetic configurations (\autoref{t0-1}). However, for monolayer $\mathrm{CrF_{1.5}I_{1.5}}$, the FM and   AF-N$\acute{e}$el orders have almost the same energy, and the difference is only 0.48 meV/$\mathrm{CrF_{1.5}I_{1.5}}$ formula. So, we  focus on the FM state of both monolayer $\mathrm{CrF_{1.5}I_{1.5}}$ and  $\mathrm{CrCl_{1.5}I_{1.5}}$ for a better comparison.
For monolayer $\mathrm{CrF_{1.5}I_{1.5}}$ ($\mathrm{CrCl_{1.5}I_{1.5}}$),  the optimized lattice constants is 6.250  (6.590) $\mathrm{{\AA}}$,  and the calculated $C_{11}$ and $C_{12}$ are  49.97 (34.88) $\mathrm{Nm^{-1}}$ and  16.34 (9.94) $\mathrm{Nm^{-1}}$, which satisfy the  Born  criteria of mechanical stability\cite{ela}. From \autoref{ph11}, it is proved  that monolayer $\mathrm{CrF_{1.5}I_{1.5}}$ ($\mathrm{CrCl_{1.5}I_{1.5}}$) is dynamically stable. The side views of crystal structures  of $\mathrm{CrX_{1.5}I_{1.5}}$ (X=F, Cl and Br) are plotted in \autoref{cccc}, and it is clearly seen that the distortions of  octahedral
environment  located by Cr atoms  become more and more severe with X from Br to Cl to F due to the more difference in atomic sizes and electronegativities of X and I atoms. It is found that an easy axis of monolayer  $\mathrm{CrCl_{1.5}I_{1.5}}$
is along the c-direction, and the corresponding MAE is  110 $\mu$eV per Cr atom. However, for monolayer $\mathrm{CrF_{1.5}I_{1.5}}$, an easy axis
is along the in-plane direction, and the MAE is up to 2151 $\mu$eV per Cr atom. Finally, the piezoelectric properties of monolayer $\mathrm{CrF_{1.5}I_{1.5}}$ and  $\mathrm{CrCl_{1.5}I_{1.5}}$ are investigated, and their  $d_{31}$ is up to  2.578 pm/V and  1.804 pm/V, respectively.
The related data are summarized in \autoref{tab0}. In fact, many PFMs can be achieved in 2D   $\mathrm{CrX_3}$ (X=F, Cl, Br and I) family by using the same design principle of  monolayer  $\mathrm{CrBr_{1.5}I_{1.5}}$, for example Janus monolayer  monolayer  $\mathrm{CrCl_{1.5}F_{1.5}}$,  $\mathrm{CrCl_{1.5}Br_{1.5}}$,   $\mathrm{CrBr_{1.5}F_{1.5}}$ and so on.

In summary, our theoretical calculations demonstrate that the
PFM  can occur in Janus $\mathrm{CrBr_{1.5}I_{1.5}}$ monolayer with dynamic, mechanical and thermal  stabilities, which possesses  a sizable MAE.
 By breaking the inversion and mirror symmetry,  both  in-plane and  out-of-plane piezoelectric polarizations can be
induced by a uniaxial in-plane  strain. Amazingly, the  out-of-plane $d_{31}$  (1.138 pm/V)  is obviously higher compared with ones of many familiar 2D materials. It is proved that strain engineering can effectively  tune piezoelectricity of  monolayer  $\mathrm{CrBr_{1.5}I_{1.5}}$.
The PAFM  can also  be realized by compressive strain, and  $d_{22}$  ($d_{31}$) is 0.677 pm/V  (0.999 pm/V) at 0.94 strain.
Finally, similar to $\mathrm{CrBr_{1.5}I_{1.5}}$, the PFM can also be achieved in the  monolayer $\mathrm{CrF_{1.5}I_{1.5}}$ and $\mathrm{CrCl_{1.5}I_{1.5}}$ with very large $d_{31}$ being  2.578 pm/V and 1.804 pm/V.
 Our works supply an experimental proposal  to achieve large out-of-plane piezoelectric response in PFMs, and hope that the work can stimulate further
experimental effort on 2D PFM.

\begin{acknowledgments}
This work is supported by Natural Science Basis Research Plan in Shaanxi Province of China  (2021JM-456). We are grateful to the Advanced Analysis and Computation Center of China University of Mining and Technology (CUMT) for the award of CPU hours and WIEN2k/VASP software to accomplish this work.
\end{acknowledgments}

\end{document}